\documentstyle[preprint,pre,aps,eqsecnum]{revtex}

\begin{document}
\draft
\draft
\title{Monte Carlo study of exact S-matrix duality\\
in non simply laced affine Toda theories
}
\author{M. Beccaria}
\address{
Dipartimento di Fisica dell'Universit\`a  and INFN\\
Piazza Torricelli 2, I-56100 Pisa, Italy\\
e-mail: beccaria@hpth4.difi.unipi.it
}
\date{\today}
\maketitle
\begin{abstract}
The $(g_2^{(1)}, d_4^{(3)})$\ pair of non simply laced affine Toda theories 
is studied from 
the point of view of non perturbative duality. 
The classical spectrum of each member is composed of two massive scalar 
particles. The exact S-matrix prediction for the dual behaviour of the 
coupling dependent mass ratio is found to be in strong agreement with Monte 
Carlo data.
\end{abstract}

\vskip 5mm

\hskip 1.15truecm keywords: Affine Toda theories, exact S-matrix, bootstrap 
principle.

\vskip 5mm

\pacs{PACS numbers: 11.10.Lm, 02.20.Fh, 11.10.Gh, 11.10.Kk}

\section{Introduction}

Affine Toda field theories are two-dimensional models described by the
Euclidean action
\begin{equation}
S = \int d^2x\left(\frac{1}{2}\partial_\mu\phi\cdot\partial_\mu\phi
+\frac{m^2}{\beta^2}\sum_{a=0}^r n_a e^{\beta \alpha^{(a)}\cdot\phi}\right).
\end{equation}
The $r$-dimensional vectors $\{\alpha^{(a)}\}$ are the simple roots of an 
affine Kac-Moody algebra~\cite{Kac} and $\{n_a\}$ are positive integers 
depending on the algebra and satisfying 
\begin{equation}
\sum_a n_a \alpha_a = 0, \qquad n_0 = 1.
\end{equation}
The field $\phi$ is a set of $r$ real scalar components. Finally, $m$ and 
$\beta$ are a mass scale parameter and the coupling constant.
The Coxeter number is the positive integer $h=n_0+\cdots +n_r$.

Under the transformation $T:\alpha\to 2\alpha/|\alpha|^2$, the lattice of the 
simple roots transforms into the lattice of another affine algebra. 
The invariant algebras are called self-dual; they belong to 
the untwisted a-d-e series $a_n^{(1)}$, $d_n^{(1)}$, $e_n^{(1)}$
and to the twisted series $a_{2n}^{(2)}$. The other algebras are the pairs
$(b_n^{(1)}, a_{2n-1}^{(2)})$, $(c_n^{(1)}, d_{2n-1}^{(2)})$,
$(g_2^{(1)}, d_4^{(3)})$, and $(f_4^{(1)}, e_6^{(2)})$; they are invariant 
under $T$.

At the classical level, affine Toda theories have no coupling; $\beta$ can be 
scaled away, the spectrum is proportional to $m$, independent of $\beta$ and 
moreover it is given by simple universal formulae in terms of  the Coxeter 
number.
The interest of the classical theory is that the field equations of motion 
admit a Lax pair and therefore
there is an infinite hierachy of conserved currents with increasing spin. 

At the quantum level, this property is inherited in the form of a factorized 
S-matrix. The dependence on $\beta$ which plays the role of Planck's constant 
becomes non trivial; on the other hand the parameter $m$ becomes unphysical 
due to renormalization effects and only mass ratios are observables.

Since the S-matrix is expected to be factorizable, its explicit form may be 
sought. One can make a guess and impose physical constraints like unitarity 
or  crossing symmetry 
and the additional bootstrap principle. In the case of the self-dual 
theories, perturbation theory suggests that the mass ratios do not 
renormalize. Indeed, the bootstrap equations close on an ansatz for the 
S-matrix based on the tree 
level spectrum and on the fusings allowed by the three-point 
couplings~\cite{Braden,Mussardo1}.

Perturbation theory and the structure of the bootstrap suggest conjectured 
expressions
for the exact $\beta$ dependence of the S-matrix which show a remarkable 
duality 
between weak and strong coupling in terms of the transformation $\beta\to 
4\pi/\beta$.

As pointed out in~\cite{Mussardo2}, for the non self dual pairs the picture is 
more complicated. Mass ratios deform already  at the lowest order of perturbation theory 
and the simplest ansatz for the S-matrix fails.

However, a non trivial solution to the bootstrap equations can be found with
the feature of predicting $\beta$ dependent mass 
ratios~\cite{Delius,CorriganA}. 
The predictions are then formally the same as in the classical 
theory, but in terms of a ``renormalized'' Coxeter number $H(\beta)$.

Again, the explicit non perturbative form of $H(\beta)$ is not known.
The simplest conjecture~\cite{Dorey}, consistent with low order perturbation 
theory~\cite{Cho} and current algebra~\cite{Kausch}
predicts a new kind of duality. Under $\beta\to 4\pi/\beta$ the S-matrices of
the pair members get exchanged. Hence, the strong coupling regime in one theory
should be given by the weak coupling regime in the other.

In~\cite{Watts} a Monte Carlo study of duality in the pair 
$(g_2^{(1)}, d_4^{(3)})$ was performed
by mean of the Metropolis algorithm. The authors determined the mass ratio in 
the $g_2^{(1)}$ theory over a wide range of couplings and they did find 
agreement 
with the duality conjecture. Specifically, they checked that the mass ratio 
in $g_2^{(1)}$ ranged 
between its classical values and the classical value of $d_4^{(3)}$ .

In this paper I carried over the above simulation on larger lattices with 
higher statistics in order to pin down the precise dependence on $\beta$. 
Moreover, I have used the Hybrid Monte Carlo algorithm~\cite{Duane}. Finally,
I have extended the simulation to the $d_4^{(3)}$ theory in order to have a 
complete picture.

The plan of the paper is the following: Section~\ref{pair} describes the pair
$(g_2^{(1)}, d_4^{(3)})$ ; Section~\ref{oneloop} shows the one loop 
deformations of the mass 
ratios; 
Section~\ref{simulation} gives some detail on the numerical simulation; 
finally,
in Section~\ref{results} the results are discussed.

\section{The dual pair ($g_2^{(1)}$, $d_4^{(3)}$)}
\label{pair}

The pair ($g_2^{(1)}$, $d_4^{(3)}$) has $r=2$ and its action is
\begin{equation}
S = \int d^2x \left\{
\frac{1}{2} \partial_\mu\phi\cdot \partial_\mu\phi
+\frac{m^2}{\beta^2}\sum_{a=0}^2
n_a \exp(\beta\ \alpha_a\cdot\phi)
\right\},\qquad \phi = (\phi_1, \phi_2).
\end{equation}
The integers $n_a$ and the simple roots are
\begin{eqnarray}
g_2^{(1)} &:& n = \{2,3,1\},\qquad 
\alpha = \left\{\left(\sqrt{2},0\right),
\left(-\frac{1}{\sqrt{2}},\frac{1}{\sqrt{6}}\right),
\left(-\frac{1}{\sqrt{2}},-\sqrt{\frac{3}{2}}\right)\right\}; \\
d_4^{(3)} &:& n = \{2,1,1\},\qquad
\alpha = \left\{\left(\sqrt{2},0\right),
\left(-\frac{3}{\sqrt{2}},\sqrt{\frac{3}{2}}\right),
\left(-\frac{1}{\sqrt{2}},-\sqrt{\frac{3}{2}}\right)\right\}.
\end{eqnarray}
The two sets of roots are related by the duality 
$\alpha\to 2\alpha/|\alpha|^2$.

The two corresponding models are very different at the tree level. 
The  explicit expansion of the mass-potential exponential 
term up to the fourth order for the $g_2^{(1)}$ model is
\begin{eqnarray}
\lefteqn{V(\phi_1,\phi_2) = m^2\left(3\,{\phi_1^2} + {\phi_2^2}\right) +} && 
\nonumber \\
&& + m^2\beta {{9\,{\phi_1^3} - 9\,\phi_1\,{\phi_2^2} - 2\,{\sqrt{3}}\,
{\phi_2^3}}\over
       {9\,{\sqrt{2}}}} + \\ 
&& + m^2\beta^2 {{27\,{\phi_1^4} + 18\,{\phi_1^2}\,{\phi_2^2} + 
       8\,{\sqrt{3}}\,\phi_1\,{\phi_2^3} + 7\,{\phi_2^4}}\over {72}} + 
O(\beta^3). \nonumber
\end{eqnarray}
For the $d_4^{(3)}$ model we utilize the tree level mass eigenstates by 
transforming
the fields
\begin{equation}
\phi\to R\phi ,\qquad R = \left(\begin{array}{cc}
\cos\theta &  \sin\theta \\
-\sin\theta & \cos\theta
\end{array}\right), \qquad \theta = \frac{5 \pi}{12}
\end{equation}
and obtain the expansion
\begin{eqnarray}
\lefteqn{V(\phi_1,\phi_2) = m^2\left( (3 - {\sqrt{3}})\,{\phi_1^2} + 
   (3+{\sqrt{3}})\,{\phi_2^2} \right) + } && \nonumber\\ 
&& +  m^2\beta\,\left( {{{\phi_1^3}}\over 2} - {{{\sqrt{3}}\,{\phi_1^3}}
\over 2} + 
     {{3\,{\phi_1^2}\,\phi_2}\over 2} - {{{\sqrt{3}}\,{\phi_1^2}\,\phi_2}
\over 2} + 
     {{3\,\phi_1\,{\phi_2^2}}\over 2} + {{{\sqrt{3}}\,\phi_1\,{\phi_2^2}}
\over 2} + 
     {{{\phi_2^3}}\over 2} + {{{\sqrt{3}}\,{\phi_2^3}}\over 2} \right)  + \\
&& +m^2\beta^2   \,\left( {{7\,{\phi_1^4}}\over 8} - 
     {{5\,{\phi_1^4}}\over {4\,{\sqrt{3}}}} + {{{\phi_1^3}\,\phi_2}\over 2} - 
     {{{\phi_1^3}\,\phi_2}\over {{\sqrt{3}}}} + {{3\,{\phi_1^2}\,{\phi_2^2}}
\over 4} + 
     {{\phi_1\,{\phi_2^3}}\over 2} + {{\phi_1\,{\phi_2^3}}\over {{\sqrt{3}}}} 
+ 
     {{7\,{\phi_2^4}}\over 8} + {{5\,{\phi_2^4}}\over {4\,{\sqrt{3}}}} 
\right) + O(\beta^3) \nonumber
\end{eqnarray}
As one can see, the sets of possible fusings are completely different and
duality is far from being obvious.
The classical mass ratios are 
\begin{equation}
\left.\frac{m_2}{m_1}\right|_{g_2^{(1)}} = \sqrt{3},\qquad
\left.\frac{m_2}{m_1}\right|_{d_4^{(3)}} = \sqrt{\frac{\sqrt{3}+1}
{\sqrt{3}-1}} = 
\frac{\sqrt{3}+1}{\sqrt{2}}
\end{equation}
which agree with the general formula in terms of the Coxeter number $h$
(6 for $g_2^{(1)}$ , 12 for $d_4^{(3)}$)
\begin{equation}
\frac{m_2}{m_1} = \frac{\sin(2\pi/h)}{\sin(\pi/h)} = 2 \cos(\pi/h)
\end{equation}
The duality conjecture states that the correct quantum ratio 
$g_2^{(1)}$ is given by substituting $h\to H(\beta)$ in the model $g_2^{(1)}$ 
and 
$h\to H(4\pi/\beta)$ in the $d_4^{(3)}$ model.
The form of $H(\beta)$ is constrained but not fixed by perturbation theory 
and the conjectured expression is
\begin{equation}
H(\beta) = 6+\frac{\beta^2/2\pi}{1+\beta^2/12\pi}.
\end{equation}
Let us clarify these statements by considering the one loop mass ratios.

\section{One loop mass ratios}
\label{oneloop}

Let us denote the three diagrams of  Figures (I-II-III) by
\begin{equation}
\Gamma^{(1)}_{abc}, \quad \Gamma^{(2)}_{abcd}, \quad \Gamma^{(3)}_{abcd}
\end{equation}
where $a$, $b$, $c$ and $d$ are particle labels in the range $\{1,2\}$. 
The mass ratio is observable since renormalization amounts to a normal 
ordering of the 
exponentials and its effect is a redefinition of the bare mass.
We must check that in a bare renormalization scheme all the divergent tadpole 
graphs cancel. 
Let us utilize dimensional regularization and let us introduce
\begin{equation}
Z_i = \int \frac{d^dp}{(2\pi)^d}\frac{1}{p^2+m_i^2} .
\end{equation} 
The pole part of $Z_i$ is mass independent, hence the cancellation is a 
matter of couplings. At the
one loop level the mixed propagator corrections are irrelevant and we can 
restrict to the diagonal
ones. Let us write the interaction lagrangian in the form
\begin{equation}
V(\phi_1,\phi_2) = \frac{1}{2}(m_1^2\phi_1^2 + m_2^2\phi_2^2) + 
V_{111}\phi_1^3 + 
V_{112}\phi_1^2\phi_2 + \cdots .
\end{equation}
Then, the corrections to the propagator of particle 1 are
\begin{eqnarray}
\Gamma^{(1)}_{111}  &=& -12\ V_{1111}\  Z_1,\nonumber\\
\Gamma^{(1)}_{112}  &=& -2\  V_{1122}\  Z_2,\nonumber\\
\Gamma^{(2)}_{1111} &=& 18\  V_{111}^2\  Z_1\  m_1^{-2}, \\
\Gamma^{(2)}_{1112} &=& 6\  V_{111}\  V_{122}\  Z_2\  m_1^{-2},\nonumber\\
\Gamma^{(2)}_{1121} &=& 2\  V_{112}^2\  Z_1\  m_2^{-2},\nonumber\\
\Gamma^{(2)}_{1122} &=& 6\  V_{112}\  V_{222}\  Z_2\  m_2^{-2}. \nonumber
\end{eqnarray}
The corrections to the propagator of particle 2 are obtained by exchanging 
the 1 and 2 labels.
If we denote the full divergent correction by
\begin{equation}
\delta m_1^2 = \Gamma^{(1)}_{111}+\Gamma^{(1)}_{112}+\Gamma^{(2)}_{1111}+
\Gamma^{(2)}_{1112}+\Gamma^{(2)}_{1121}+\Gamma^{(2)}_{1122} 
\end{equation}
then the desired cancellation is equivalent to the condition
\begin{equation}
\frac{\delta m_1^2}{m_1^2}=\frac{\delta m_2^2}{m_2^2}
\end{equation}
which is indeed satisfied by the couplings of the two theories which can be 
read in the 
expansions of the previous section.

Besides the consistency check, let us turn to the mass ratio deformation. 
At one loop, we must determine the quantity
\begin{equation}
\delta\ \frac{m_1^2}{m_2^2} = \frac{m_1^2}{m_2^2} \left(
\frac{\delta m_1^2}{m_1^2}-\frac{\delta m_2^2}{m_2^2}
\right).
\end{equation}
Let us introduce the finite integral
\begin{equation}
Z_{ij}(p^2) = \int \frac{d^2 q}{(2\pi)^2} 
\frac{1}{(q^2+m_i^2)((q+p)^2+m_j^2)}.
\end{equation}
Then the finite contributions to the propagators of particle 1 are
\begin{eqnarray}
\Gamma^{(3)}_{1111} &=& 18 V_{111}^2 Z_{11}(p^2),\nonumber\\
\Gamma^{(3)}_{1112} &=& 4 V_{112}^2 Z_{12}(p^2),\\
\Gamma^{(3)}_{1122} &=& 2 V_{122}^2 Z_{22}(p^2).\nonumber
\end{eqnarray}
Evaluation on the tree mass shell gives
\begin{equation}
-\delta m_1^2 = 18 V_{111}^2 Z_{11}(-m_1^2) + 4 V_{112}^2 Z_{12}(-m_1^2)
+2 V_{122}^2 Z_{22}(-m_1^2)
\end{equation}
with analogous expressions for the particle 2.
We need only the following particular values
\begin{eqnarray}
Z_{ii}(-m_i^2) &=& \frac{1}{4\sqrt{3}}\frac{1}{m_i^2}, \\
Z_{ij}(-m_i^2) &=& \frac{1}{2\pi\sqrt{m_j^2(m_j^2-4m_i^2)}}\mbox{ArcTanh}
\sqrt{\frac{m_j^2-4m_i^2}{m_j^2}}
\end{eqnarray}
and the final result is 
\begin{equation}
g_2^{(1)}: \qquad \delta\  \frac{m_1^2}{m_2^2} = \frac{1}{12\sqrt{3}} \beta^2 
+ O(\beta^4); \qquad
d_4^{(3)}: \qquad \delta\  \frac{m_1^2}{m_2^2} = -\frac{1}{16} \beta^2 + 
O(\beta^4).
\end{equation}
The renormalized Coxeter number is thus
\begin{equation}
g_2^{(1)}: \qquad H(\beta) = 6 + \frac{\beta^2}{2\pi} + O(\beta^4); \qquad
d_4^{(3)}: \qquad H(\beta) = 12 - \frac{9\beta^2}{2\pi} + O(\beta^4) 
\end{equation}
and a consistent, simple and natural conjecture is
\begin{equation}
g_2^{(1)}: \qquad H(\beta) = H_0(\beta); \qquad
d_4^{(3)}: \qquad H(\beta) = H_0(4\pi/\beta) 
\end{equation}
where 
\begin{equation}
H_0(\beta) = 6 + \frac{\beta^2/2\pi}{1+\beta^2/12\pi} \qquad 6<H_0<12.
\end{equation}
The result for $g_2^{(1)}$ is in agreement with that quoted by~\cite{Delius}. 
The result for $d_4^{(3)}$ gives perturbative support to the duality 
conjecture 
$\beta\to4\pi/\beta$.
We remark that the discrepancy with~\cite{Cho} is due to the fact that they 
use the 
form of $H_0$ which is correct for the simply laced models.

\section{Details of the simulation}
\label{simulation}

The lattice action for the pair ($g_2^{(1)}$, $d_4^{(3)}$) expressed in terms 
of pure numbers is
\begin{equation}
S_{\rm Toda} = \sum_{n\in\rm sites}\left\{
\frac{1}{2}\sum_{\mu=1,2} 
(\phi_{n+\mu}-\phi_n)^2 + \frac{m^2}{\beta^2}\sum_{a=1}^3
n_a \exp(\beta\ \alpha_a\cdot\phi)
\right\},\qquad \phi = (\phi_1, \phi_2).
\end{equation}
I have simulated the Toda theory with the Hybrid Monte Carlo algorithm 
(see~\cite{Duane} for
the details). 
Let us consider the extended action
\begin{eqnarray}
S &=& S_p + S_{\rm Toda}, \\
S_p &=& \frac{1}{2} \sum_n \pi_n\cdot \pi_n, \qquad \pi = (\pi_1, \pi_2).
\end{eqnarray}
The free parameter of the algorithm are $N_{hmc}$ and $\epsilon$. The first
is the number of molecular dynamics steps. The second is the time step in the
integration of the equations of motion
\begin{eqnarray}
\dot{\phi}_n &=& \pi_n, \\
\dot{\pi}_n  &=& \sum_\mu (\phi_{n+\mu}-2\phi_n + \phi_{n-\mu}) -
\frac{m^2}{\beta} \sum_a n_a \alpha_a \exp(\beta\alpha_a\cdot\phi_n).
\end{eqnarray}         
The vacuum expectation value of the field is a non physical quantity. 
However, it is interesting to measure it since it is an indicator of
thermalization and also because it is in a sense a dynamic minimum of the 
Toda potential.

Mass ratios can be determined by studying the eigenvalues of the two point 
function
\begin{equation}
\langle 0 | \Phi_i(0)\Phi_j(\tau)| 0 \rangle - \langle 0 | \Phi_i| 0 \rangle 
\langle 0 | \Phi_j| 0 \rangle
\end{equation}  
where $t$ is the lattice time ranging from $0$ to $T$ and the wall field 
$\Phi_i(t)$ is obtained by averaging $\phi$ over space.

\section{Results}
\label{results}

I have used a $80^2$ lattice for all $\beta$s because the correlation lenght 
may be adjusted
by varying $m$. The continuum mass ratio is independent 
of the bare mass $m$. However, on a finite lattice, it must be chosen
in order to have correlation lenghts large with respect to one lattice 
spacing and small compared to the lattice size. This is the correct 
procedure which minimizes discretization and finite size corrections.
Thanks to the work of~\cite{Watts} I had good values in the case of the
$g_2^{(1)}$ theory. In the other model, I started with the same values of $m$
adjusting them for some $\beta$. 

I utilized different measurement of the wall-wall two-point
function for each bin in the separation $\tau$. This is necessary in 
order to avoid strong correlation between data.

Table~\ref{table1}  shows the Hybrid Monte Carlo parameters which we
found to be optimal for each couple $(\beta,m)$. The time step $\epsilon$
must be reduced almost exponentially as $\beta$ is increased. This is 
reasonable since at larger $\beta$ the potential profile becomes steeper.

Table~\ref{table2} shows the measure of $\langle 0 | 
\phi_i | 0 \rangle$ which can be useful as a check 
of the code and which is needed in order to subtract the two-point function.

Table~\ref{table3} shows the two lattice masses, their ratio and the
conjectured prediction.

Finally, tables~\ref{table4}, \ref{table5}, \ref{table6} show the same results 
in the case of the $d_4^{(3)}$ model.

Figures I-II-III show the self energy diagrams which are needed in order to 
compute the one loop mass ratio deformations.

Figure IV shows a summary plot of the measured mass ratios in the two
models together with the conjectured ones and the asymptotic values holding 
in the classical limit.

\section{Conclusions}

In this paper I have investigated numerically the conjectured duality 
in the pair $(g_2^{(1)}, d_4^{(3)})$ of non simply laced affine Toda theories.
I have shown
that the $\beta$ dependence of the mass ratios in $g_2^{(1)}$ does follows the 
behaviour conjectured in~\cite{Watts} and that the data of $d_4^{(3)}$
agree with the $\beta\to4\pi/\beta$ duality.

As in the case of more realistic field theories like QCD, the numerical 
approach could be useful in studying other interesting features of quantum 
Toda theories. 
For instance, one could try to find direct evidence of the boundary bound 
states which appears when the theory is restricted to a 
half-line~\cite{CorriganB}; work is in progress on this topic. 
Moreover, it could be valuable a non perturbative study of the solitons 
which appear at imaginary $\beta$, and which suggests that a unitary theory 
can ultimately be found by restricting the state space of the 
hamiltonian~\cite{Hollowood};
their stability is indeed still questionable~\cite{Sasaki}.

\section{Acknoledgements}
I gratefully acknowledge G.~M.~T.~Watts and R.~A.~Weston for 
useful suggestions and interest.

\newpage

\section*{Captions}

\noindent{\bf Fig. I-II-III  :} self energy one loop diagrams.

\noindent{\bf Fig. IV        :} summary of the numerical results.

\begin{table}
\caption{$g_2^{(1)}$: HMC parameters.}
\label{table1}
\vskip 1truecm
\begin{center}
\begin{tabular}{|ccc|}
$\beta$ & $N_{hmc}$ & $\epsilon$ \\
\tableline
1.0     & 10         & 0.1   \\
2.0     & 10         & 0.08  \\
3.5     & 10         & 0.07  \\
5.0     & 10         & 0.06  \\
10.0    & 5          & 0.05  \\
20.0    & 10         & 0.025 \\
\end{tabular}
\end{center}
\end{table}

\begin{table}
\caption{$g_2^{(1)}$ bare mass and dynamical minimum.}
\label{table2}
\vskip 1truecm
\begin{center}
\begin{tabular}{|cccc|}
$\beta$ & $m$  & $\langle\phi_1\rangle$ & $\langle\phi_2\rangle$ \\
\tableline
1.0     & 0.1   & -0.0982(6)  & 0.225(1)  \\
2.0     & 0.1   & -0.1542(4)  & 0.3768(8) \\
3.5     & 0.05  & -0.2073(5)  & 0.5300(9) \\
5.0     & 0.01  & -0.2752(4)  & 0.6866(8) \\
10.0    & 5E-5  & -0.3396(7)  & 0.843(1)  \\
20.0    & 5E-7  & -0.2362(4)  & 0.7023(7) \\
\end{tabular}
\end{center}
\end{table}

\begin{table}
\caption{$g_2^{(1)}$: mass ratio.}
\label{table3}
\vskip 1truecm
\begin{center}
\begin{tabular}{|ccccc|}
$\beta$ & $m_1$ & $m_2$ & $R$ & $R^*$ \\
\tableline
1.0     & 0.1595(2) & 0.27839(5) & 1.745(2) & 1.74509\\
2.0     & 0.2133(2) & 0.3784(1)  & 1.775(2) & 1.77605\\ 
3.5     & 0.2262(5) & 0.41192(5) & 1.821(5) & 1.82579\\ 
5.0     & 0.1615(4) & 0.3004(1)  & 1.861(5) & 1.86150\\ 
10.0    & 0.1379(3) & 0.2630(2)  & 1.908(5) & 1.90870\\ 
20.0    & 0.2896(3) & 0.5576(1)  & 1.926(3) & 1.92562\\ 
\end{tabular}
\end{center}
\end{table}


\begin{table}
\caption{$d_4^{(3)}$: HMC parameters.}
\label{table4}
\vskip 1truecm
\begin{center}
\begin{tabular}{|ccc|}
$\beta$ & $N_{hmc}$ & $\epsilon$ \\
\tableline
1.0     & 10        & 0.07   \\ 
2.0     & 10        & 0.08   \\ 
3.5     & 10        & 0.05   \\ 
3.5     & 10        & 0.06   \\ 
5.0     & 10        & 0.05   \\ 
10.0    & 10        & 0.02   \\ 
10.0    & 10        & 0.03   \\ 
20.0    & 10        & 0.0125 \\ 
\end{tabular}
\end{center}
\end{table}

\begin{table}
\caption{$d_4^{(3)}$ bare mass and dynamical minimum.}
\label{table5}
\vskip 1truecm
\begin{center}
\begin{tabular}{|cccc|}
$\beta$ & $m$  & $\langle\phi_1\rangle$ & $\langle\phi_2\rangle$ \\
1.0     & 0.1   &  0.1530(5)  & -0.1930(9) \\ 			
2.0     & 0.1   &  0.2014(3)  & -0.2210(6) \\
3.5     & 0.05  &  0.2305(2)  & -0.2174(4) \\
3.5     & 0.01  &  0.3163(5)  & -0.3687(9) \\		
5.0     & 0.01  &  0.2701(3)  & -0.2552(5) \\
10.0    & 5E-5  &  0.3061(3)  & -0.2729(6) \\
10.0    & 1E-7  &  0.4410(8)  & -0.504(1)  \\
20.0    & 5E-7  &  0.2488(2)  & -0.1489(4) \\	 
\end{tabular}
\end{center}
\end{table}

\begin{table}
\caption{$d_4^{(3)}$: mass ratio.}
\label{table6}
\vskip 1truecm
\begin{center}
\begin{tabular}{|ccccc|}
$\beta$ & $m_1$ & $m_2$ & $R$ & $R^*$ \\
\tableline
1.0     & 0.2166(1)  & 0.4149(1)  & 1.916(2)  & 1.91665 \\
2.0     & 0.3363(2)  & 0.6323(1)  & 1.880(1)  & 1.88120 \\
3.5     & 0.3427(1)  & 0.6280(1)  & 1.832(1)  & 1.82840 \\
3.5     & 0.2150(1)  & 0.3931(3)  & 1.829(1)  & 1.82840 \\
5.0     & 0.3850(2)  & 0.6903(4)  & 1.793(2)  & 1.79398 \\
10.0    & 0.3506(6)  & 0.6155(3)  & 1.756(4)  & 1.75193 \\
10.0    & 0.1201(3)  & 0.2106(1)  & 1.753(6)  & 1.75193 \\
20.0    & 0.5727(6)  & 0.9966(6)  & 1.740(2)  & 1.73740 \\
\end{tabular}
\end{center}
\end{table}

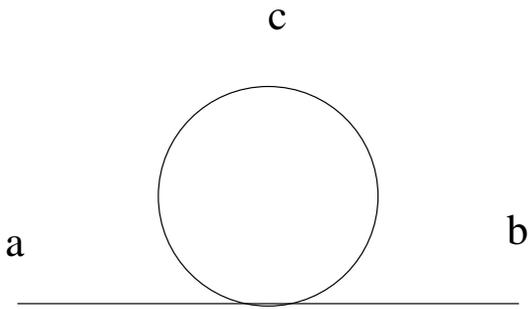
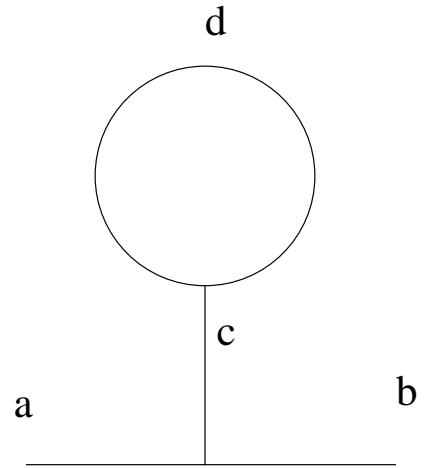
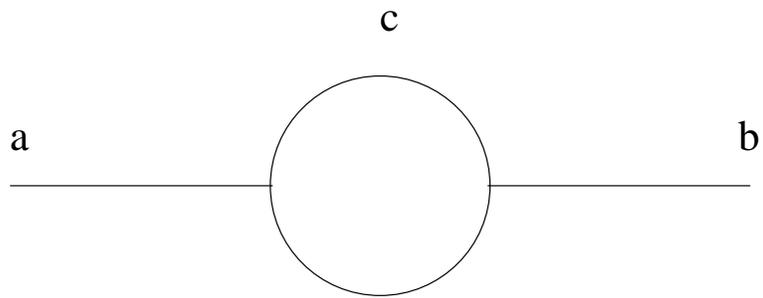
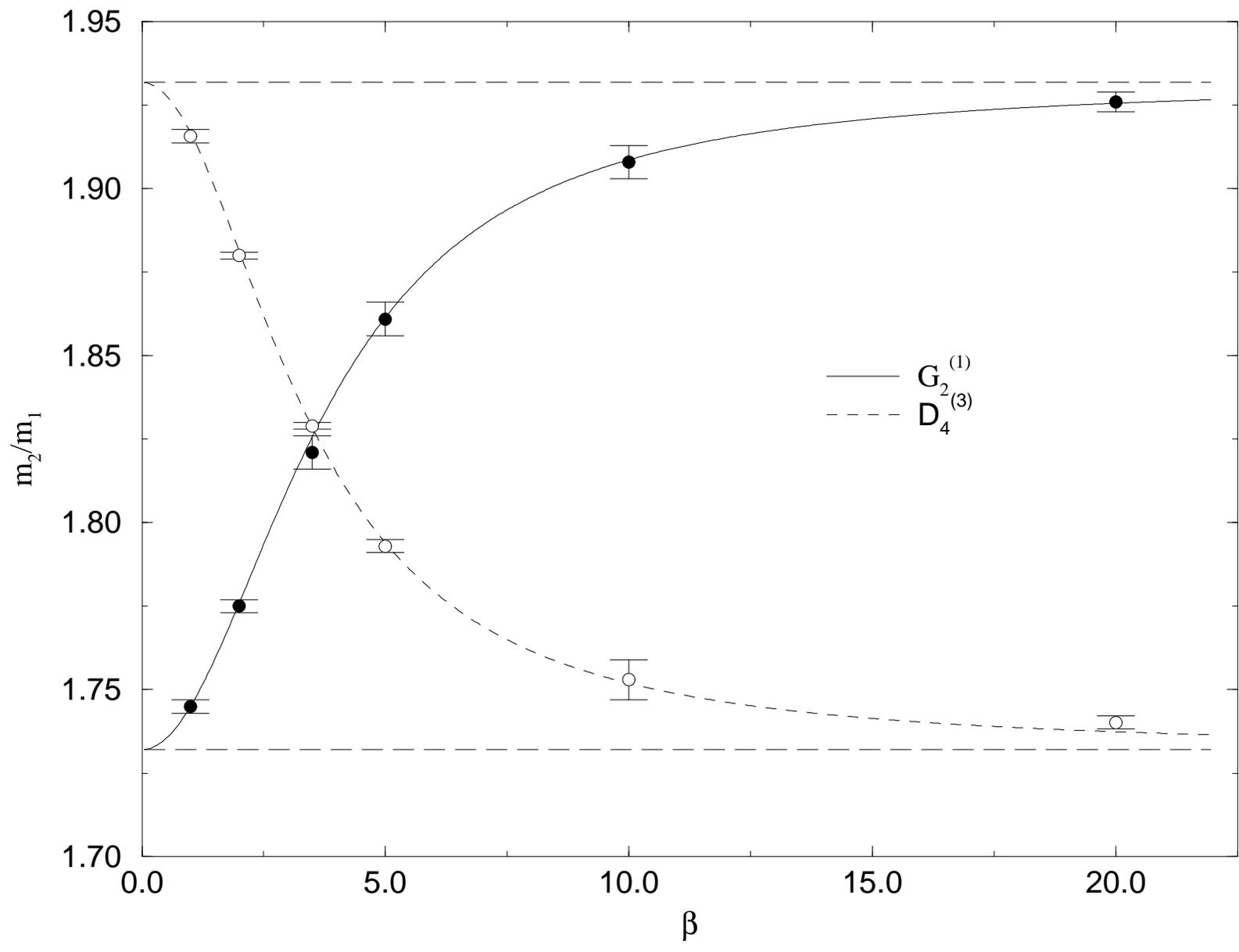
\end{document}